\documentclass[10pt,letterpaper]{article}
\usepackage[top=0.85in,left=2.75in,footskip=0.75in]{geometry}
 
\usepackage{amsmath,amssymb}
\usepackage{xurl}
\usepackage{hyperref}
\usepackage{multirow}
\usepackage{changepage}

\usepackage[utf8x]{inputenc}

\usepackage{textcomp,marvosym}

\usepackage{cite}

\usepackage{nameref,hyperref}

\usepackage{microtype}
\DisableLigatures[f]{encoding = *, family = * }

\usepackage[table]{xcolor}

\usepackage{array}

\usepackage{color}
\newcolumntype{+}{!{\vrule width 2pt}}

\newlength\savedwidth

\raggedright
\usepackage[aboveskip=1pt,labelfont=bf,labelsep=period,justification=raggedright,singlelinecheck=off]{caption}

\makeatletter
\renewcommand{\@biblabel}[1]{\quad#1.}
\makeatother

\usepackage{lastpage,fancyhdr,graphicx}
\usepackage{epstopdf}
\fancyheadoffset[L]{2.25in}
\fancyfootoffset[L]{2.25in}
\begin{document}
\vspace*{0.2in}

\begin{flushleft}
{\Large
\textbf\newline{Digital Information Seeking and Sharing Behavior During the First Wave of the COVID-19 Pandemic} 
}
\newline
\\
Mehk Fatima\textsuperscript{1},
Aimal Rextin\textsuperscript{2},
Mehwish Nasim* \textsuperscript{3},
Osman Yusuf \textsuperscript{4}
\\
\bigskip
\textbf{1} Asthma and Allergy Institute Pakistan and Department of Computer Science \& Information Technology, University of Lahore, Gujrat Campus, Gujrat, Pakistan
\\
\textbf{2}  Asthma and Allergy Institute Pakistan and Department of Computer Science, COMSATS University Islamabad, Islamabad, Pakistan
\\
\textbf{3} Flinders University, Australia.
\\
\textbf{4}  Asthma and Allergy Institute Pakistan
\\
\bigskip

%
%





*mehwish.nasim@flinders.edu.au

\end{flushleft}
\section*{Abstract}
People turn to search engines and social media to seek information during population-level events, such as during civil unrest, disease outbreaks, fires, or flood. They also tend to participate in discussions and disseminate information and opinions via social media forums, and smartphone messaging applications. COVID-19 pandemic was not any different. However, the proper medical awareness and correct information dissemination is critical during a pandemic. An unprecedented amount of internet traffic related to this topic was generated during the start of the pandemic all over the world.  
   
   In this work, we have analysed the electronic data generated by users from Pakistan on Google Search Engine and WhatsApp to understand their information-seeking behavior during the first wave of the pandemic.  The paper aims at analysing how the Pakistani public developed their understanding about the disease, (its origin, cures, and preventive measures to name a few) by analysing digital data. We found that the public actively searched and discussed information at the start of the pandemic. However, their interest waned with time and was reinvigorated only when something novel or shocking seemed to have occurred.    Understanding this \emph{information seeking behavior} will allow corrective actions to be taken by health policymakers to better inform the public for possible future waves of a pandemic through electronic media, as well as and the social media companies and search engines to address misinformation among the users in the emergent markets.



\section*{Introduction}
first wave of COVID-19 pandemic tested the health care systems and economies around the world. A key point noted in Pakistan was the lack of understanding about the causes, symptoms, and preventive measures of COVID-19. That frequently led to lax attitude towards social distancing protocols\cite{wolf2020awareness} or widespread adoption of pseudo-medicinal remedies that are known be ineffective or to have dangerous side effects \cite{Gastro2011}.

The World Health Organization \cite{Review2009}, after the H1N1 pandemic considered reliable communication as one of the most pivotal components to fight a pandemic. In the current pandemic, social media played a key role in the spread of both information and misinformation about the pandemic. For example, in Spain, WhatsApp was the second most used channel to acquire information about the pandemic \cite{moreno2020}. Since the information on social media is not always reliable hence it is not surprising that in extreme cases this has led to people using hazardous chemicals to wash their vegetables or clean their throats even in developed countries like the US\cite{Joseph2020}. Two malarial prophylaxis drugs: Chloroquine and Hydroxychloroquine, got massive public attention in the last week of March 2020 after the endorsement of multiple high-profile figures including President Donald Trump, resulting in online searching of chloroquine increasing by 442\% and hydroxychloroquine by 1389\% in America \cite{Liu2020}.

In October 2020, Gallup Pakistan released a survey report that described the public’s behavior on COVID-19, just before the increase in the number of new infections during second wave in Pakistan. According to that report, almost 75\% of the population thought that COVID-19 was under control and the need for continued precautions was no longer necessary. Whereas, nearly 70\% of the public thought that the threat of the SARS-CoV-2 virus was exaggerated, 46\% consider COVID-19 as a foreign conspiracy, and 45\% public thought that it was a laboratory-made virus. Hence, according to the survey, the public underestimated the threat posed by this disease and many even considered it unreal \cite{gallup2020}.

Conveying correct medical information to the public is extremely important in a country like Pakistan with limited medical resources, and the high prevalence of pseudo-medicine and quackery in order to ensure that the information seekers have been provided with correct and reliable information. The aim of this paper is to understand how the Pakistani public developed its understanding of this disease, its origin, cures, and preventive measures.  Understanding this \emph{information seeking behavior} will allow corrective actions to be taken by health policymakers to better inform the public for possible future waves of this pandemic, especially through electronic channels.

\begin{figure}[!h]
\frame{\includegraphics[width=\linewidth]{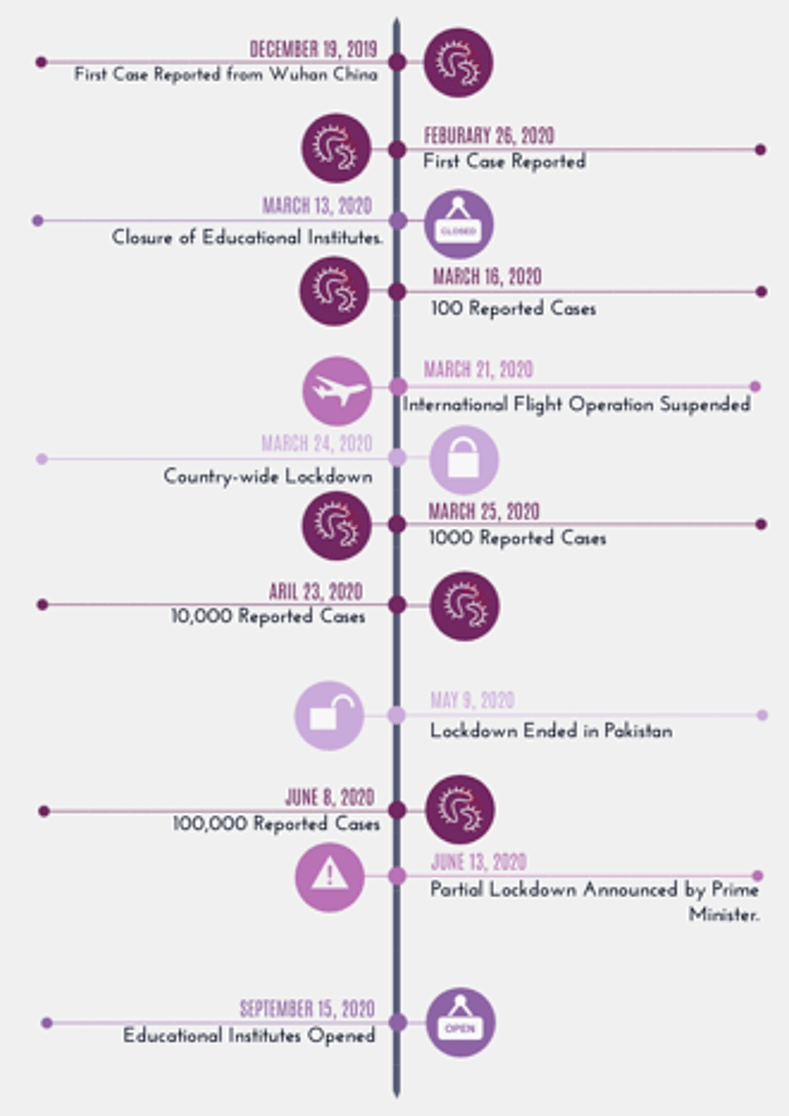}}
\caption{{Timeline of major events during first wave of COVID-19 pandemic in Pakistan.}}
\label{fig:events}
\end{figure}

We now review how the first wave of the pandemic progressed as it is likely to influence how people seek information about it online.   Figure \ref{fig:events}  summarises major events during the first wave from a local perspective. 

\subsection*{Organization}
The paper is organized as follows: Section $2$ summarizes studies relevant to this research. In Section $3$ we will give an overview of our method including which sources of digital information we will use and our strategy to analyse this data. In Section $4$ we will present the results from our first source of digital information, Google Trends, while in Section $5$ we present the results from analysing the $2{nd}$ digital data source, WhatsApp. Section $6$ concludes the paper. 
\section*{Related Work} \label{sec:RL}

  Due to the sheer volume of available online information, it is very difficult for a common person to differentiate between information and misinformation \cite{Pennycook2020}. Ideally, a rational and careful selection of their information diet \cite{Rizi2020} is needed which is frequently not easy. Hence it is no surprise that according to a WHO report published in 2011, effective communication is one of the biggest challenges that need to be handled in a global pandemic \cite{WHO2011}.

The public acquired the needed information about the SARS-CoV-2 virus from various formal and informal sources and then applied various strategies to make sense of information \cite{Genuis2012} \cite{Genuis2006}. These information sense-making techniques included the expertise of the source, the tone and style of language, the sponsor of the site and whether there is consensus between multiple sources \cite{Sun2019}. There are various instruments available to assess the reliability of online health information such as Health on Net \cite{Boyer1998}, the JAMA benchmark \cite{Silberg1997} and DISCERN \cite{Rees2002}.  However, surprisingly most health information websites lack these instruments and hence making it difficult for a layperson to assess the reliability of available health information \cite{Baltazar2020}.

Still, the internet is used extensively for health-related information, for example, in one study 56.1\% of endocrinology patients said that they seek health information online, with 77.1\% believing the information to be of high-quality \cite{Kryriacou2019}. In another study, senior adults (mean age 77 years) internet was trusted more than television or radio for health-related information but less than their social relations especially those connected to healthcare \cite{Chaudhuri2013}, whereas younger adults and adolescents \cite{Gray2005} are still more likely to trust the health information available on the internet more than older people \cite{Miller2011,McPherson2013}. A study conducted in Australia reported that myths about COVID-19 were more believed by lesser young educated men and those with non-English speaking background \cite{bonner2020}.

Online search trends can be very helpful in digital surveillance and prediction of an infectious disease. Relative search volume regarding COVID-19 increased during the early period of the pandemic and there was a positive correlation between daily new cases and relative search volume \cite{D2020}\cite{Walker2020}. A study found a strong correlation between COVID-19 related Google trends and daily new cases in the US, with $R$ value around $0.80$ \cite{Kurian2020}. Similarly, a sharp hike in Google trends happened in searches related to COVID-19 after the detection of the first case in Taiwan 
\cite{Husnayain2020}.  This strong correlation between the daily confirmed cases and related Google search trends worldwide can be used to predict the trends of outbreak \cite{Kurian2020} \cite{Ayyoubzadeh2020}. The rapid increase in web searches on COVID-19 and related topics also created an infodemic like situation and caused the worldwide spread of misinformation on disease \cite{Rovetta2020,Grau2020,Rathore2020}. Effective strategies are needed by governments and public health organizations to better manage such infodemic and strengthen the public awareness on the outbreak \cite{D2020}. Educating the public to use websites of official public health forums can be helpful in this regard \cite{Garcia2020}.

Social media played a key role in propagating health-related misconceptions and poses a big challenge to practitioners and policymakers \cite{gupta2020information}. The main reason for this challenge is that many people are not clear about the relationship between science, policy-making and media \cite{authortatize}, and they tend to rely more on nonscientific but more definitive advice.  For example, misinformation circulating during the 2014 Ebola outbreak challenged the efforts of health workers to control the epidemic \cite{Chou2018}. Even in countries like Germany, Italy, US and UK, social media movements incited people to resist getting measles vaccination \cite{Datta2018},\cite{Antonietta2017}. Similarly, another study analyzed $2691$ tweets about the treatments and preventive measures of gynecologic cancer and found $30\%$ of them to contain misinformation \cite{Chen2018}. While another study found that $40\%$ of links shared on health-related forums contained fake news and  were shared more than $450,000$ times between 2012-2017 \cite{Waszak2018}.

Psychological and cognitive biases greatly influence how we react in a pandemic. People anticipated that the SARS-CoV2 virus cases would grow linearly, and they underestimated the possibility of an exponential growth \cite{denworth2020}. This lack of understanding and failure of public health messaging had disastrous results even in developed countries \cite{shah2020}. Conspiracy theories and myths pose another serious threat to public health and affect the behavioral responses of people \cite{moyer2019} that can dangerously affect the situation in a pandemic. Pseudo-medicinal information and conspiracy theories about COVID-19 circulated through social media traveled faster than the virus itself \cite{Depoux2020}. Many websites containing unproven claims about COVID-19 are widely visited by the public. They also shared on social media sometimes due to naivety and surprisingly sometimes intentionally to share inaccurate information \cite{Pennycook2020}. This lack of accurate health information on COVID-19 also severely affected the psychological condition of the general public during the pandemic \cite{Wang2020}.

Several attempts are made to track the online health-related pseudo-medicinal information. Ghenai and Mejova built a tool to track misinformation about health concerns on Twitter. They collected more than $13$ million tweets during the Zika outbreak in 2016 and applied automated machine learning techniques to identify rumor-carrying tweets \cite{Ghenai2017}. In a similar study, researchers tried to track rumors about vaccines on social networks \cite{Kostkova2017}. Another study presented a customizable interactive web-based framework that integrates various heterogeneous health related data streams including tweets, news, reports etc. They specifically focus on HPV vaccine scare-related data and visualize the results in the form of several graphs using this framework \cite{Kostkova2016}. Another study based on user-centric modeling presented a classifier to identify users who are more prone to spread pseudo-medicinal information about fake cures of cancer on Twitter. They characterized the users based on several user attributes, linguistic styles, sentiments, and post timings \cite{Ghenai2018}.

Google search trends, search queries, and social media debates can be used to analyze the public interests on a specific topic. This data can be specifically helpful during international crises and epidemics. A few studies tried to predict the spread of epidemics in a specific geographical location by performing analysis on search engine queries and Google trends in that region \cite{Ginsberg2009,Polgreen2008,Teng2017}. Search query analysis also showed that the public immediately started searching about the pandemic but they started searching for prevention and protection e.g.  social distancing in the initial stages of the COVID-19 pandemic \cite{Bento2020}.  Similarly, studies indicated that social media can also be a useful way for early detection of epidemics \cite{Kim2014} \cite{Thapen2016}.

\section*{Method}\label{sec:method}
In this section we will present our methodology to understand the digital information seeking behaviour of Pakistani users during the first wave pandemic of COVID-19. Although digital information can take many forms like text, images, and videos, we focused on analysing text data due to ease of availability and simplicity of analysis.
It is popular to study the spread information or misinformation in text based digital data on online social networking sites such as Twitter\cite{sharma2020coronavirus}\cite{kuvsen2018politics}. However, we took an alternative route and chose the following two other sources of digital information:
\begin{enumerate}
\item We studied how information was searched on \textbf{Google search engine} for COVID-19 related searches through Google Trends. We focused more on search queries that are known to be pseudo-medicinal information on COVID-19. We then compared these search trends with real world circumstances like changing government SOPS, the number of cases, and number of deaths etc. This will help us understand the information trajectory in Pakistan during the current COVID-19 crises

\item WhatsApp is a popular way of communication in Pakistan. According to a mobile ranking forum, WhatsApp is the second most used mobile application in the country.  Owing to the popularity of this communication medium, we identified a public \textbf{WhatsApp group}  focused on COVID-19. We then exported the conversations in the form of textual data from that group. In an automated way, we replaced all names and phone numbers with unique identifiers and we then  stored the data as a .CSV file with the following columns:  Time and Date, User Identifier, Text Message.  We then analysed it to better understand the mood and the content of these messages. 

\end{enumerate}

To better understand the vast volume of digital text data, we first needed to define keywords so the data can be mined more easily. It was obvious to use corona, covid etc as these keywords but it was not clear which pseudo-medicinal treatment we should use as keywords. Next we will discuss how we made our list of pseudo-medicinal treatments. 

We shared with the participant's important information about the study. There were three stages in which this study handled data about humans. The first stage was when we aimed to identify keywords for the Google Trends and public WhatsApp group data. A written consent form and important information about the study were shared with all participants who filled our online questionnaire. Google Trends is publicly available and is aggregated trends of millions of search queries hence no verbal or written consent was taken. Similarly, the WhatsApp group was a public group, and data only anonymized data was downloaded for analysis, so no consent was needed. The Departmental Ethics Committee of COMSATS University, Islamabad reviewed the study protocol and confirmed that this study adheres to required ethical standards. The approval letter was numbered on 19-05-2020/CS/ETHICS.

Data of public WhatsApp groups has been used in social science research \cite{Bursztyn2019}, \cite{Garimella2018}.  Bursztyn et. al suggest that users must be aware that their data will be shared with other members once they participate in a group, public or not \cite{Bursztyn2019}. 
When we started this work, we informed the ethics committee in detail about the methodology and they agreed that only informing the group members will suffice, as there is no objection from the platform's end on saving WhatsApp messages by the admin of the group.  We communicated with the group admin for permission to use their data, the admin then communicated with the group members, and received no objection from any group member. The privacy policy for WhatsApp groups states that a user share their messages and profile information with other members of the group (https://www.whatsapp.com/legal/)  and they can also save and email upto 10,000 messages to anyone (https://faq.whatsapp.com/en/android/23756533/).  The data was then anonymized using a script written in the language R.

\subsection*{Pseudo-Medicinal Treatments}
We observed that misinformation about the treatment of COVID-19 were circulating on social media during the early months of COVID-19 in Pakistan. The authors of this papers observed the following non-scientific treatments of COVID-19  were popular on social media platforms during the first wave of the pandemic in Pakistan:
\begin{enumerate}
\item \textbf{Herbs and Spices} like garlic, ginger tea, lemon tea, olive leaf and senna makki (a laxative). Albeit some of these are harmless but senna makki, being a laxative, can cause dehydration and in extreme cases deaths in patients. 
\item \textbf{Homeopathic drugs} like Arsenicum album received publicity as a powerful immunity booster and were thought to help in preventing COVID-19. While the drug by itself may not have any serious side effects but people consider themselves immune to COVID-19 and as consequence took lesser preventive measures. 
\item \textbf{Medicines} like Chloroquine (also marketed with brand name Resochin) and Hydroxychloroquine are antimalarial drugs that gained a lot of attention on social and traditional media without any solid scientific backing and sometimes can be  dangerous. Further this medicine is also used for rheumatoid arthritis patients and over-the counter availability of this drug created a shortage in the market for the patient in need. 
\item \textbf{Convalescent Plasma} therapy was a popular treatment in Pakistan \cite{khan2020} and worldwide including U.S. despite many concerns on its effectiveness and possible adverse effects in some cases \cite{Malani2020}.
\end{enumerate}

The above list of unproven treatments of COVID-19 can act as keywords for our textual analysis. However, it is not clear if the general Pakistani public was aware of these terms and hence used it for in their online search queries and conversations. So we tried to confirm the suitability of using these keywords through a small non-representative sample of users.   We electronically  sent our consent form and  questionnaire to potential participants and then proceeded with only those who indicated their consent electronically. The questionnaire lists  misinformation  on social media (as listed above) as  keywords and asked the participant to tick the boxes which they recognized.  The targeted population of this survey was the general Pakistani public and we tried to get the representation from different areas of the country. A total of 40 participants (18 Males and 22 Females) filled this online survey. The participants were geographically distributed in 19 locations around Pakistan from residents of small towns to those living in big cities. 

\begin{figure}[!h]
\begin{center}
    
\frame{\includegraphics[width=\linewidth]{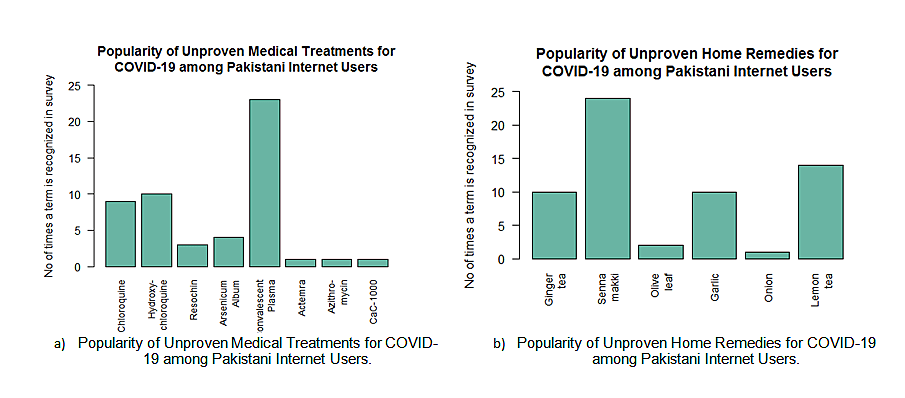}}
  \caption{Popularity of Unproven treatments identified on different social and electronic media forums among Pakistani internet users. We can see that  the most popular home treatment was Senna Makki, followed by Lemon Tea, Ginger Tea, and Garlic. On the other hand, plasma therapy was a very popular medicinal treatment followed by Hydroxychloroquine and Chloroquine }
 \label{fig:misinfo}
\end{center}
\end{figure}

All home remedies identified by our research team were validated through this survey. As shown in Figure~\ref{fig:misinfo} (a), the most popular home treatment was Senna Makki, followed by Lemon Tea, Ginger Tea, and Garlic. Figure~\ref{fig:misinfo} (b) shows the popularity of unproven medical treatments for COVID-19 in Pakistan, which shows that a lot of interest was shown by internet users in plasma therapy as a possible treatment for COVID-19 plasma therapy followed by Hydroxychloroquine and Chloroquine.

\section*{Analysis $1$: Search Trends During COVID-19 Pandemic}\label{sec:search-Analysis}
We performed a systematic analysis to evaluate the relationship between Google search trends on various non-scientific treatments on pseudo-medicinal information and the changing situation of COVID-19  pandemic in Pakistan. We also tried to map the government initiatives to deal with the pandemic and major events such as the first COVID-19 case and the first death. We also evaluated their effects on Google trends on popular pseudo-medicinal information topics in the country.

\subsection*{Dataset}
\emph{Google Trends} represents the popularity of a specific search term on Google during a specific duration.  For a specific search term, Google Trends shows the daily relative popularity of the search term during this duration. It returns a number $n$ ranging between $0$ and $100$, where $n=100$ on the day when the search term was most popular and $n=0$ when it was least popular. This allows the analysis of search interests of users in specific regions as well as around the globe. It also provides the comparison of search trends on multiple search terms by similarly normalizing them between $0$ and $100$. This allows us to compare the relative popularity of multiple search terms, giving us insight to public interests and concerns at a specific time.  

\begin{figure}[!h]
\begin{center}
    
\frame{\includegraphics[width=\linewidth]{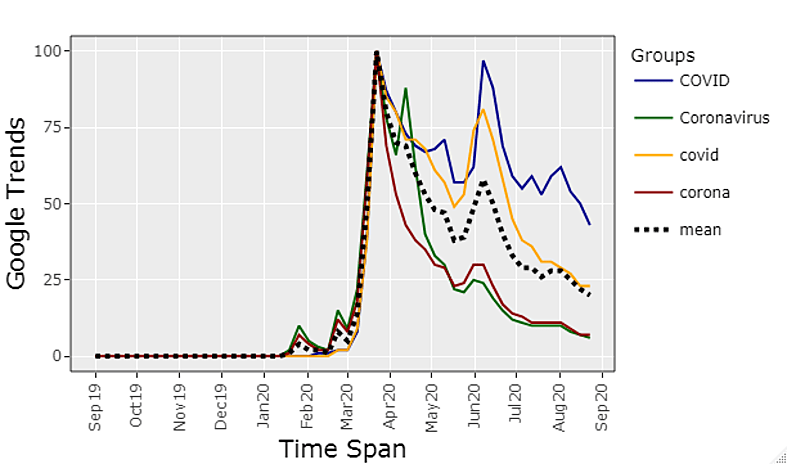}}
	\caption{{Comparison of Google Trends on Coronavirus search terms during pandemic in Pakistan. The dashed line shows the mean popularity of COVID-19 related searches.}}
\label{fig:terms}

\end{center}
\end{figure}

We first performed a comparative analysis of Google Trends on different terms that can be alternatively used for searching details about COVID-19 in Pakistan. We choose $4$ generic terms that are commonly used in Pakistan: COVID-19, Coronavirus, Covid, and Corona. As shown in Figure \ref{fig:terms}, all these terms started to appear in Google Trends during the 3rd week of January 2020 and they all reached their peak in 3rd Week of March 2020. A sudden rise appeared in search trends during the second Week of March as the government implemented various spread control measures including nationwide lockdown. We also calculated their mean popularity on Google Trends and in the remainder of this paper, we will only use the mean of the various terms used to search for Coronavirus  as shown in Figure \ref{fig:terms}.

\begin{figure}[!h]
\begin{center}
 \frame{\includegraphics[width=\linewidth]{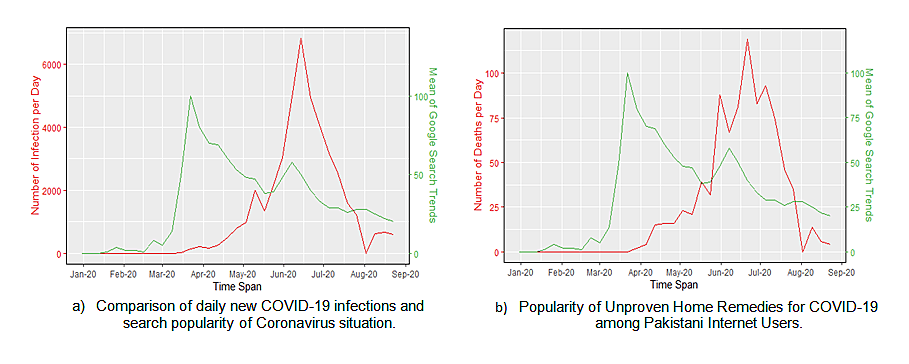}} 
  \label{fig:coronaVsDeaths}
  \caption{A comparison of how internet search statistics about COVID-19  changed with changing number of infections and deaths.}
  
 \label{fig:effect_pandemic_searches}
 
\end{center}
\end{figure}

We also checked whether the quick spread of COVID-19 in the region and the rising number of infections and deaths resulted in increased searches in COVID-19 by the public. Here our assumption is that the increased number of searches indicate increased public concern about the pandemic.  We explored this  by comparing the day to day statistical data of COVID-19 cases in Pakistan with the popularity of COVID-19 searches using data from Google Trends. We computed the correlation of search popularity of COVID-19 searches with daily new infections and found that it has a value of just $0.27$. Similarly, the correlation of popularity of COVID-19 searches with daily deaths came out to be $0.23$. We can see that these  are very small values, indicating that the search popularity was insignificantly influenced by the spread of the pandemic.  Figure \ref{fig:effect_pandemic_searches} (a) shows how search popularity change with rising number of new infections and similarly Figure \ref{fig:effect_pandemic_searches} (b) shows how search popularity change with daily deaths. It is surprising as well concerning that people seem to be searching very little even at the peak of the pandemic and it seems that Google searches about COVID-19 was fuelled more because of the initial total lack of knowledge about the virus and the resulting disease.

\subsection*{Search Interest Regarding Treatment and Prevention of COVID-19}

\begin{figure} [!h]
\frame{\includegraphics[width=\linewidth]{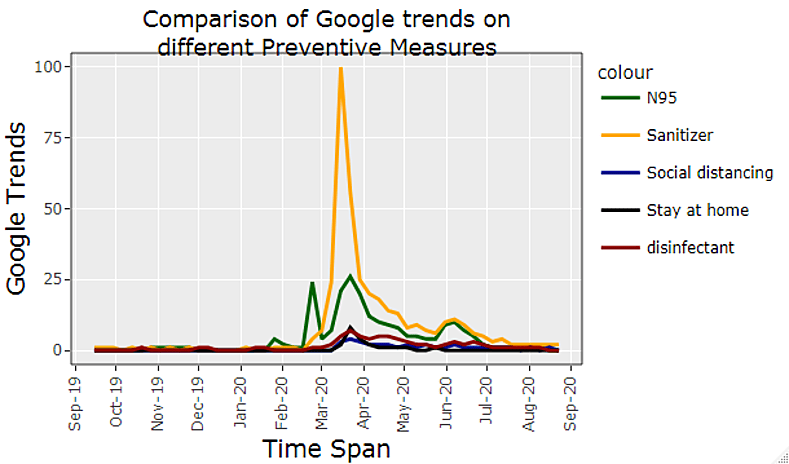}}
	\caption{Comparison of Google Trends on Preventive Measures during Pandemic in Pakistan.}
\label{fig5}
\end{figure}

We also performed a comparative analysis of Google trends in Pakistan on different preventive measures suggested by WHO for COVID-19. We choose 5 popular terms that were commonly used in Pakistan: N95 mask, Sanitizer, Social distancing, stay at home, disinfectant. Figure~\ref{fig5} shows the comparison and it is clear that “Sanitizer” is clearly the most popular of these terms during the first wave of the pandemic in Pakistan. We used the top trending term “Sanitizer” for further analysis to evaluate the relationship between these trends and peaks of trends on different pseudo-medicinal information search terms.

\begin{figure}[!h]
\frame{\includegraphics[width=\linewidth]{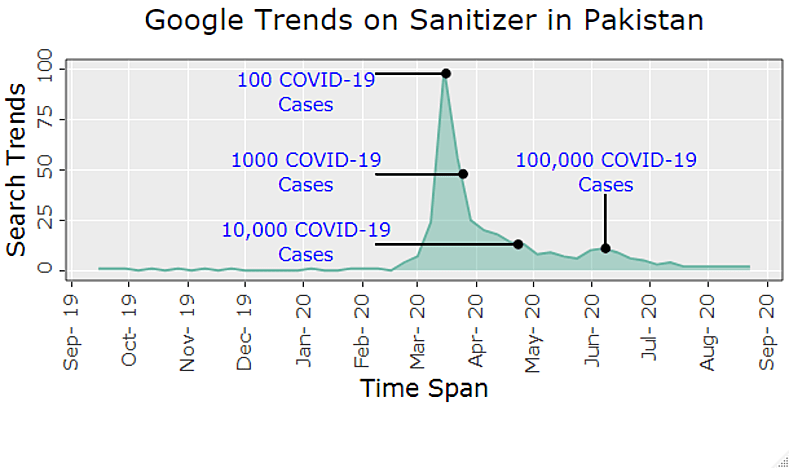}}
	\caption{Comparison of Google Trends on Sanitizer during March-August 2020.}
\label{fig:sanitizer}
\end{figure}

Figure~\ref{fig:sanitizer} shows the pattern of how the popularity of the search term  \textit{sanitizer} varies over time. We can see that this search term experienced a sudden rise in popularity immediately after the first patient was detected, this was  when the whole country was experiencing previously unknown levels of fear and uncertainty. At this time, many  items essential for the pandemic like sanitizers, masks, and other hygiene related items experienced a sudden increase in demand.  This was probably when the public was searching if sanitizers were available online or if they are available at cheaper price. However, very soon private businesses started to fill this newly created demand and as a result  the search term "sanitizer" quickly dropped in popularity However, it still remained more popular than pre-pandemic time.

\begin{figure} [!h]
\centering
\frame{\includegraphics[width=\linewidth]{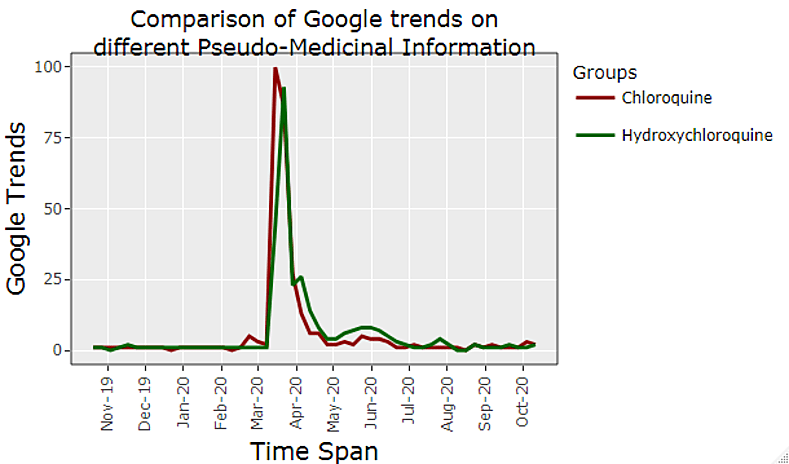}}
	\caption{Comparison of Google Trends on Chloroquine and Hydroxychloroquine in Pakistan.  }
\label{fig7}
\end{figure}

We also investigated the Google Trends to check the popularity of “Chloroquine” and “Hydroxychloroquine” in Pakistan after Trump's endorsement of these drugs and found a pattern very similar to that in the US. As presented in Figure~\ref{fig7}, there was a sudden hike in Google searches on both medicinal  terms in the last week of March, 2020 and they remain popular till 3rd week of April, 2020. This indicates that the backing of a popular personality drives the public interests and increases the trustworthiness of a piece of information. 

In summary, the results presented suggests that people in Pakistan actively search COVID-19 related information during the early stages of the first wave of the outbreak but later the public interest seemed to have waned.

\section*{Analysis $2$: WhatsApp Public Group Data}\label{sec:WhatsApp}
Now,  we present an analysis of data from a public WhatsApp group and also explore the popularity of the pseudo-medicinal information in the discussions during the first wave of COVID-19.

At the start of the pandemic, the general public used various social media forums to share information about the newly emerging disease. WhatsApp was among the most popular forums used during the pandemic to spread the information and pseudo-medicinal information [4]. Hence, we thought it would be beneficial to try to get a broad understanding of the discussions on WhatsApp about COVID-19. We explored many public WhatsApp groups for discussions about it COVID-19 and its pseudo-medicinal keywords mentioned in previous sections. But, we found that the keywords were mentioned insignificant number of times. Therefore, we decided to focus  on only one public WhatsApp group specifically created in response to the COVID-19 pandemic called \texttt{Understanding COVID-19}. The group was created to understand the causes and potential cures of this novel disease and its members actively discussed various issues about the pandemic, especially in the start of the pandemic when  limited information about this disease was available. 

This particular group was created in the middle of March 2020, when the number of daily infections and deaths started to grow.  This group  has $53$ members,  all of whom are Pakistani residents and most of them  live in different major cities of Pakistan. Moreover, almost all members were literate and with basic healthcare knowledge and some of them were even medical doctors. We reached this conclusion as many members of the group introduced themselves when the group was created. Even those who didn’t introduce themselves had very good English and their tone and language indicated that they are well educated.

The complete record of this WhatsApp group was exported till the 3rd week of August as a CSV file for text analysis in \texttt{R}.  Since most of the discussion on this group was in English with Urdu being used only occasionally, so the text analysis was done only on text messages in English. As a prepossessing step, we performed the following data cleaning steps,  so the data could be be more meaningful.

\begin{enumerate}
    \item All characters in Urdu or any other language were removed and we only considered ASCII characters for our text analysis. Hence all emojis were also ignored. Note if a message was written in Roman Urdu, then it was included in our analysis. 
    The national language of Pakistan when  transliterated  in Roman script is called as \emph{Roman Urdu}  \cite{bilal2017roman}. This is a common practice in Pakistan due to difficulties in entering text on Urdu keyboards. .
    \item Any multimedia files shared on the group were ignored. 
    \item Web links shared on the group were also ignored, however, if a user had copy pasted text from a website than those were part of the analysis. 
\end{enumerate}

\begin{figure}[!h]
\frame{\includegraphics[width=\linewidth]{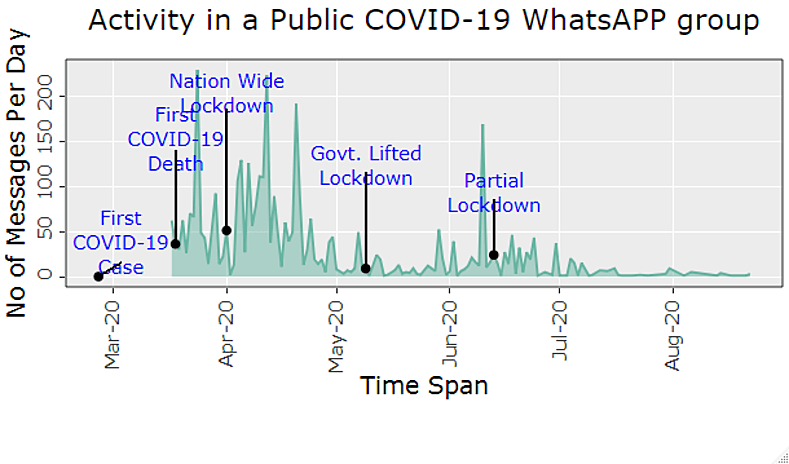}}
	\caption{{A representation  of message frequency per day in COVID-19 public group with important events in Pakistan. An increase in the number of messages in a day can be seen after two important events, the day the first patient died and the day with the national lock.}}
\label{fig:whatsapp-time-count}
\end{figure}

The time series visualized in Figure \ref{fig:whatsapp-time-count} represents the frequency of messages per day and the points indicate important events during the first wave of the COVID-19 pandemic in Pakistan. We can see in Figure~\ref{fig:whatsapp-time-count}, the group remains mostly active between March and July 2020, the period when COVID-19 cases were at its peak. An increase in the number of messages in a day can be seen after two important events, the day the first patient died and the day when the national lockdown was initiated. 

\begin{figure}[!h]
    \begin{center}
        
   \frame{\includegraphics[width=\linewidth]{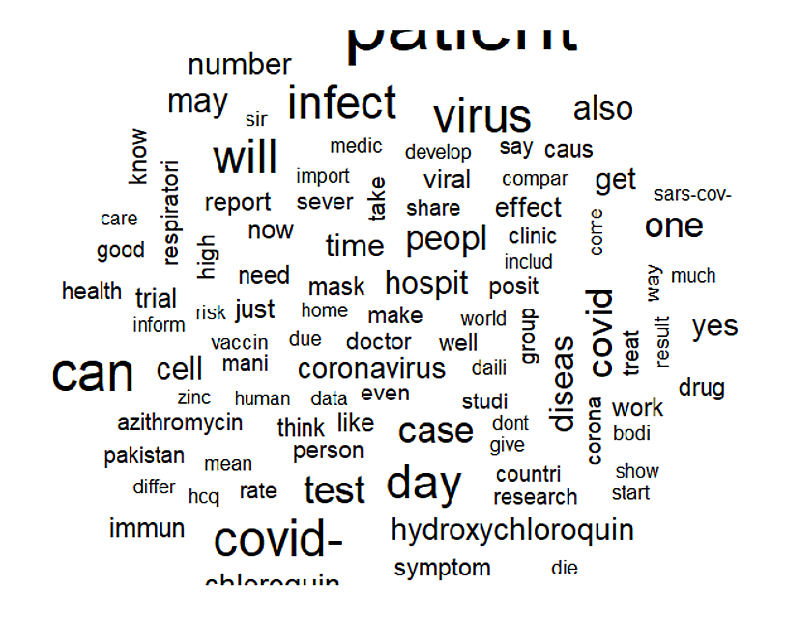}}
    \caption{The word cloud generated from the text from the WhatsApp group showing the most commonly used words in their discussions. We can observe that there is a lot of concern in its participants including discussions on patients, clinics, hospitals, medics, and also various possible treatments.}
    \label{fig:cloud}
    
    \end{center}
\end{figure}

We then wanted to get an idea of the discussion in this WhatsApp group. We plotted a  word cloud as it gives a good idea of the most frequent words used and hence the discussions between the group members. We performed the below pre-processing steps so our plot can be more meaningful.
\begin{itemize}
    \item We converted all text to lower case.
    \item  We removed all punctuation marks.
    \item We then performed stemming i.e. reduced all words to their root form.
\end{itemize}
These pre-processing steps were completed through the \texttt{stm} package of \texttt{R}. We then plotted the word cloud using the same package as shown in Figure \ref{fig:cloud}. The word cloud generated from the text from the WhatsApp group shows the most commonly used words in their discussions. We can observe by looking at the most common words that there is a lot of concern in its participants including discussions on patients, clinics, hospitals, medics, and also various possible treatments.

\begin{figure}
\begin{center}
 \frame{\includegraphics[width=\linewidth]{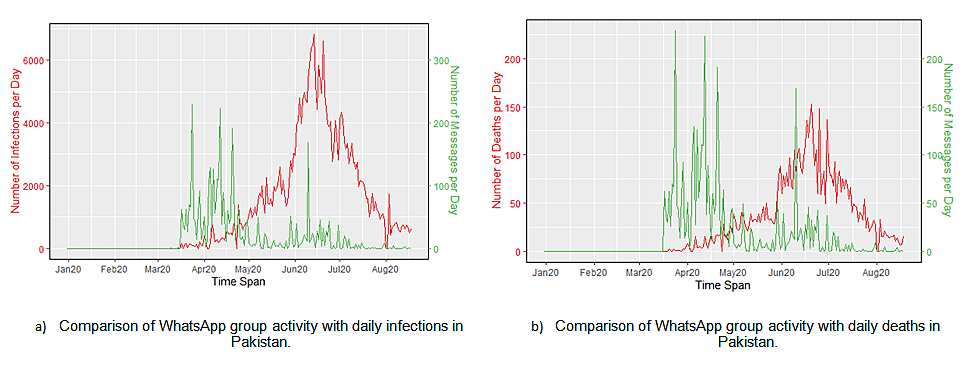}}
  \caption{Comparison of WhatsApp group activity with daily reported COVID-19 numbers in Pakistan. Please note that the peaks in (b) look larger because the number of deaths are significantly less than the number of infections.The $4$ peaks in the WhatsApp messages count seem to influenced more by coverage in  traditional news media and initial fear among the public.}
  \label{fig:whatsapp-vs-pandemic}
\end{center}
 
\end{figure}

We also wanted to check whether the increasing number of COVID-19 infections and deaths also resulted in increased discussion on the WhatsApp group, indicating  increased public interest and concern about the pandemic. For this purpose, we first computed the correlation, which came out to be $0.0006$ for daily new infections  and $-0.0135$ for the daily new deaths respectively. It indicates that the discussion was not correlated at all with how fast the pandemic is spreading, in fact in case of new deaths it seems that in some cases higher number of deaths seem to have resulted in lower number of messages on the WhatsApp group.  We hence compared how the time series of daily number of messages exchanged on the group changes the time series for the daily number of infections and daily number of deaths respectively. Figure\ref{fig:whatsapp-vs-pandemic} (a) compares the number of messages per day with the number of new COVID-19 number of cases per day in Pakistan, while Figure \ref{fig:whatsapp-vs-pandemic} (b)compares it with daily deaths due to COVID-19. This patterns shows the concern of the public at the very start of the pandemic and as the number of infections and deaths increased. However, very soon  with the increasing number of infections and deaths in Pakistan. It can be seen  from both figures that  that the group was more active in earlier phases of the COVID-19 (i.e., March-July 2020) and later the public seemed to be less concerned about the pandemic and possible dangers from it. The $4$ peaks in the WhatsApp messages count seem to influenced more by coverage in  traditional news media and initial fear among the public. More specifically, the first peak was around the time soon after the government implement country wide complete lockdown, the second peak came because there was a lot discussion about possible treatments and sharing individual experiences; the 3rd peak came because of there was a lot of discussion about psychological issues resulting from the pandemic and sharing views on how to cope with this new reality; while the 4th peak in mid June was because this was when the total number of infections in Pakistan crossed $100,000$. 

\begin{figure}[!h]
\centering
\frame{\includegraphics[width=\linewidth]{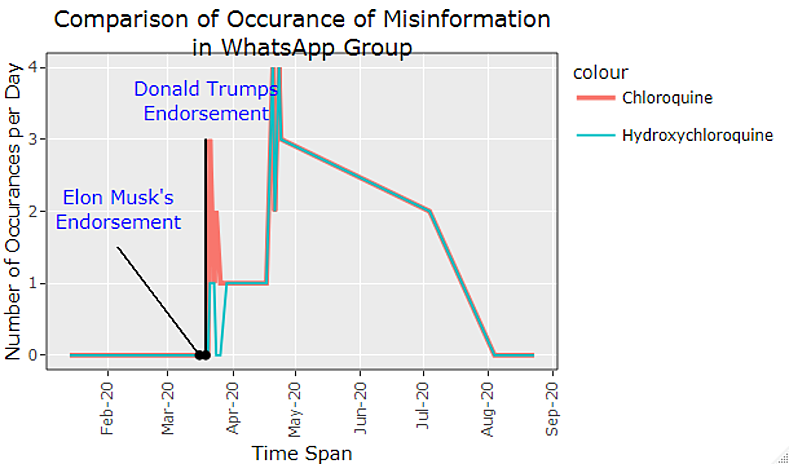}}
	\caption{Chloroquine and Hydroxychloroquine started being  discussed immediately after the endorsement of the high profile personalities and remained in discussion  till July 2020.}
\label{fig11}
\end{figure}

We also explored communication related to the two anti-malarial drugs Chloroquine and Hydroxychloroquine, and found that both started to be discussed immediately after the endorsement of President Trump and Elon Musk. They remain in regular discussion till July 2020 as can be seen in Figure \ref{fig11}. This was  despite of  the fatality reported on March 22, 2020 \cite{Liu2020}.

Lastly, we applied thematic analysis to extract main topics in the WhatsApp group.  We started by applying an algorithmic technique called Latent Dirichlet Allocation but did not find the results to be satisfactory. As a result, we followed the 6-step process suggested by Braun and Clarke for thematic analysis
\cite{braun2006using}. We note here that \emph{thematic analysis} studies are suited for studies that are of exploratory
nature and to generate hypotheses that can be later tested from a representative sample. After initial scanning of the messages for getting a high-level idea of about the issues being discussed in this group, the first author went through each message and identified 11 categories labeling each message accordingly. This labeling was
done to uncover the repeated patterns of behavior i.e. \emph{themes} in the data. The second author then reviewed this categorization by going through the file and reviewing  $40\%$ messages and their associated labels and suggested merging some categories and also recategorized some other messages. After this recategorization, we ended up with $6$ categorizes. The categories and their brief explanation is given below:

\begin{enumerate}
    \item \textbf{General Conversation:}  General conversations messages such as hellos, goodbyes, greetings, etc .
    \item \textbf{Virus Information:}  Messages that sought and shared information about the SARS-CoV-2 virus, its origin, how it spreads etc.
    \item \textbf{COVID-19 Information:}  In this category, the participants discussed the disease its risks, possible treatments and shared information that can be helpful to the patients of COVID-19.
    \item \textbf{Fake Information:}  In this category, the participants discussed various fake or dubious treatments like plasma therapy.   We also included messages that discussed or shared various conspiracy theories about the origin of the virus or the nature of the disease.  
    \item \textbf{Social Distancing:}  Messages that discussed the importance of social distancing and masks etc. or expressed concern about the non-compliance of social distancing protocols.
    \item \textbf{Anxiety/Stress:}  Messages that discussed high levels of anxiety and stress due to the pandemic were categorized into this category. This category also included suggestions intended to help cope with these issues, including those of religious nature.
   
\end{enumerate}

\begin{figure*}[!h]
\centering
\frame{\includegraphics[width=\textwidth]{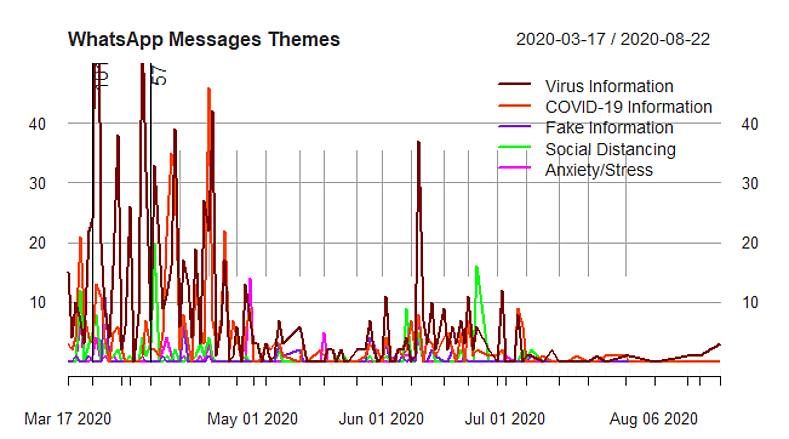}}
	\caption{Themes observed in the discussion on the public WhatsApp group. We can see that there was a lot of interest at the start of the pandemic and gradually interest reduced. Interestingly, we observed that all themes tend to be discussed more or less together with most of the discussion being focused on either information about the virus or information about the disease.  }
\label{fig:whatsapp_themes}
\end{figure*}

   The variation of these themes with time is shown in Figure \ref{fig:whatsapp_themes}. We did not include the category of general conversation in the interest of clarity of the figure. Moreover, the plot does not show two very high peaks, but their values are written where the peak is truncated. We can see from the figure that discussions in this group started as soon as it was created and the initial discussions revolved around basic information about the virus, its origin, the disease, and social distancing other preventive measures. In about a week's time, the discussion also started to have significant messages on fake information and also anxiety/stress issues. A large number of messages were exchanged for about 2 months till about May 2020 and when the volume of messages in all themes dropped. However, all themes started being discussed again with renewed interest in the middle of June, but the level of interest was lesser than the initial weeks of the pandemic and it dropped to insignificant levels within a few weeks. Recall that mid-June was the time when the total number of infections in Pakistan crossed $100,000$. Interestingly, we observed that all themes tend to be discussed more or less together with most of the discussion being focused on either information about the virus or information about the disease.

\begin{figure}[!h]
    \centering
    \frame{\includegraphics[width=\linewidth]{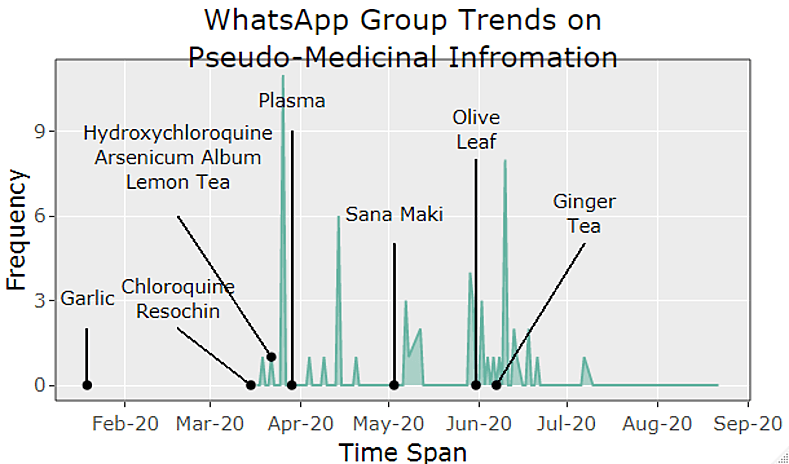}}
    \caption{The trend of how various fake information was discussed on the WhatsApp group. In this figure, the days when the search peak on Google Trends was observed is labeled. Interestingly, we can see that the search peak corresponds very near to a peak in the WhatsApp discussion about some fake information. }
    \label{fig:fakeInfo}
\end{figure}

We plotted the trend of how discussions on fake and pseudo-medicinal remedies identified in Section \ref{sec:method} varied with time.  We can see from Figure \ref{fig:fakeInfo} that this discussion seems to be bursty and the peaks seem to fall very close to the time with the peak of pseudo-medicinal treatment on Google Trends. This indicates that both the WhatsApp discussion as well as the search peak seem to be related.
\section*{Conclusion}\label{sec:conclusions}
In this paper, we investigated the information-seeking behaviour of the Pakistani public during the first wave of the COVID-19 pandemic in Pakistan. We decided to focus on online resources and social media as they were a major source of health-related information during the COVID-19 pandemic. The major contributions of this paper are the following:
\begin{enumerate}
   
    \item We investigated how the public searched the web for various COVID-19 related information as the pandemic progressed. We used data from Google Trends for this purpose. Interestingly, our analysis seems to indicate that although the number of infections and number of deaths due to COVID-19 was increasing, the general public was searching lesser for COVID-19 related information. Analysis of search trend of COVID-19 related treatments and prevention also indicated that search volume seems to be influenced more by external factors like Donald Trump and Elon Musk endorsing a drug; sanitizer not being available in the market, or the psychological effect and media coverage when the number of patients crossed $100,000$.
    \item For a high-level semantic understanding about the information sought during the first wave, we analyzed data from a public WhatsApp group that was created to share information about COVID-19. Similar to search trends, the group members shared more information during the early weeks of the pandemic and gradually the number of messages decreased despite the pandemic becoming more widespread. Like search trends, discussion in this group seemed to be sparked by external factors and associated media coverage. For example, a sudden increase in the volume of messages was observed when the COVID-19 patient died; when government-imposed countrywide complete lockdown; and when the number of infected persons crossed the $100,000$ threshold. 
    
\end{enumerate} 

The results discussed in Section \ref{sec:search-Analysis} and Section \ref{sec:WhatsApp} seem to indicate that the public was very concerned at the start of the pandemic, however with time their level of concern gradually reduced. Their level of concern and interest then seem to rise occasionally when an event that they perceived as occurred.  This seems to indicate habituation, here \emph{habituation} is psychological behavior found in all living things and it informally means the reduction of a particular response after repeated exposure of the same stimulus \cite{gluck2008learning}. A number of studies have been conducted to establish the causality of repeated exposure of the same stimulus and resulting decrease in response\cite{davis1981habituation,malcuit1996habituation}. In case our paper, the stimulus is the daily pandemic-related news that everyone heard or read at various places and the response was their level of concern and resulting online behavior that we tried to observe. Habituation is also found to be \emph{stimulus specific} \cite{thompson1966habituation}, hence the response can be returned to previous levels when an individual who is already habituated to one stimulus is presented with a novel stimulus \cite{dewsbury1981effects}.

\begin{figure} [!h]
     \centering
         
    \frame{\includegraphics[width=\linewidth]{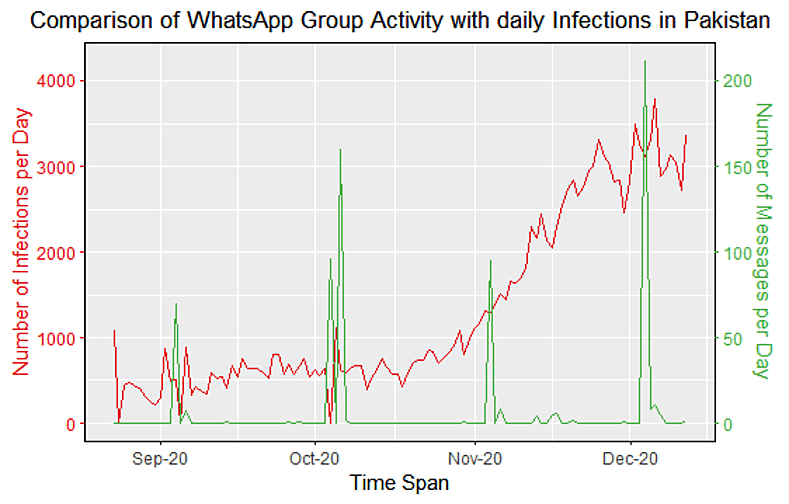}}
            \caption{Number of Messages on the Whatsapp group in the second wave compared to number of infections.}
        \label{fig:whatsapp_2ndwave}
\end{figure}

 We hence conjecture that in the case of the COVID-19 pandemic, the public habituated after daily exposure of warnings and worrying news on traditional media as well as on social media. They were, however, `dishabituated' when the stimulus changed,  for example when news of the number of infections crossing $100,000$. We check this conjecture by plotting the number of messages exchanged on the WhatsApp group in the second wave as shown in Figure \ref{fig:whatsapp_2ndwave}. Note this plot only shows till December 2020 as the WhatsApp group was deactivated soon after it. We can see that very few messages were exchanged, however, the number of messages on the group experienced sudden and short peaks. On further investigating,  they all seem to be a result of something unusual or different, explained as follows :
\begin{enumerate}
    \item \textbf{First Peak:} (September 04, 2020)  At this time, the public was concerned about the likely effect of government's decision to open schools.
    \item \textbf{Second Peak:} (October 06, 2020) A famous newspaper Dawn publishes a gloomy article\cite{dawn2} about second wave and how deadly it can be. There was a lot of discussion about this article.  
    \item \textbf{Third Peak: } (November 04, 2020) There was discussion about the news\cite{dawn3} that the government ruling out lockdown again.  
    \item \textbf{Fourth Peak:}  (December 04, 2020)  Clinical trials of the COVID-19 vaccine starts in Pakistan and it was discussed a lot in the group.
\end{enumerate}

\begin{figure}[!h]
  \frame{\includegraphics[width=\linewidth]{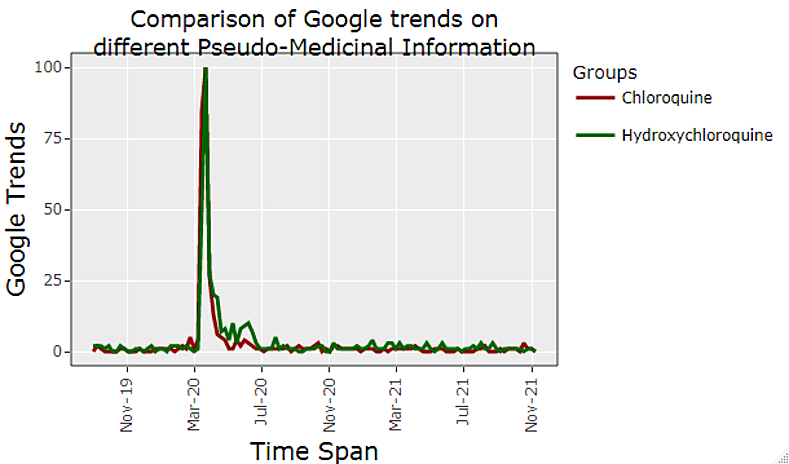}}

\caption{Figure showing Google Trends data of keywords selected in Section \ref{sec:search-Analysis} were only very high from March 2020 till about July 2020 probably due to initial scare and lack of information.}
  \label{fig:till_now}
\end{figure}

This same pattern can be seen for the search results data from Google Trends as can be seen in Figure \ref{fig:till_now} that the keywords chosen in Section \ref{sec:search-Analysis} were very high from March 2020 till about July 2020 probably due to initial scare and confusion. After the initial scare at the height of the first wave  they  dropped to almost pre-pandemic times. Hence supporting our conjecture that public's habituation to COVID-19 after an initial scare and lack of information. 

However, more work is needed to confirm whether there is a causal relationship between online activity and external factors. It also needs to be investigated how governments and health agencies should communicate with the general public so habituation can either be avoided or reduced in future pandemics. Most importantly there is a need for social media and communication applications companies (such as Facebook) and search engines (such as Google), to control the spread of misinformation, especially for the emergent user communities.

%
%
%

\bibliography{bilio}

\end{document}